\documentclass[11pt]{article}
\usepackage{graphicx,amsmath,amssymb,mathrsfs}
\usepackage[a4paper,  margin= 1 in]{geometry}
\usepackage[utf8]{inputenc}
\usepackage[english]{babel}
\usepackage[utf8]{inputenc}
\usepackage{amsmath}
\usepackage{float}
\usepackage{amsfonts}
\usepackage{amssymb} 
\usepackage{authblk}
\usepackage{natbib}
\usepackage[none]{hyphenat}
\usepackage{setspace}
\setcitestyle{authoryear,open={(},close={)}}

\begin{document}

\title{\textbf{ Observation on F.W.E.R. and F.D.R. for correlated normal   }}

\author[1]{Nabaneet Das}
\author[2]{Subir K. Bhandari}
\affil[1]{Indian Statistical Institute,Kolkata}
\affil[2]{Indian Statistical Institute,Kolkata}

\maketitle

\maketitle

\begin{abstract}
In this paper, we have attempted to study the behaviour of the family wise error rate (FWER) for Bonferroni's procedure and false discovery rate (FDR) of the Benjamini-Hodgeberg procedure for simultaneous testing problem with equicorrelated normal observations. By simulation study, we have shown that F.W.E.R. is a concave function for small no. of hypotheses and asymptotically becomes a convex function of the correlation. The plots of F.W.E.R. and F.D.R. confirms that if non-negative correlation is present, then these procedures control the type-I error rate at a much smaller rate than the desired level of significance. This confirms the conservative nature of these popular methods when correlation is present and provides a scope for
improvement in power by appropriate adjustment for correlation. 

\end{abstract}

\section{Introduction}
\label{intro}
Recently multiple hypothesis testing under dependence has gained importance due to its increased relevance in modern scientific investigations. Although many efforts have been made to generalize the existing methods under dependence (\cite{yekutieli1999resampling},
\cite{benjamini2001control}, 
\cite{sarkar2002some},\cite{sarkar2008methods},
\cite{efron2007correlation},\cite{efron2012large} etc.), very few literature is available which explicates the effect of dependence on the existing methods. \\

Correlation is one of the most important measure of dependence in the study of normal random variables. Although it is not an exhaustive measure of dependence, it might be a good starting point in order to study how dependence among observations affect the existing multiple testing algorithms. \cite{efron2010correlated} in his study of empirical Bayes methods, has shown that, the correlation penalty depends
on the root mean square (RMS) of correlations. An excellent review of the whole literature can be found in \cite{efron2012large}. \\

However, we wanted to focus on the effect of correlation in a different manner. The traditional family-wise error rate (F.W.E.R.) and the false discovery rate (F.D.R.) are the most commonly used measures of type-I error rates of a multiple testing procedure. Bonferroni's method, Holm's procedure provides strong control over F.W.E.R. and the Benjamini-Hodgeberg algorithm provides control over F.D.R. (Under independence and some positive dependence structures mentioned in \cite{sarkar2008methods}). When correlation is present, these methods shows some undesirable characteristics (Being too conservative when positive correlation is present in some cases). We have considered equicorrelated normal observations for our study and we are mostly interested in the behaviour of the type-I error rates here. If we can get an idea about how conservative they become under positive correlation, then it will provide a direction towards the required modification of the existing methods. Some distribution-free bound on F.W.E.R. can be found using Chebyshev-type inequalities mentioned in \cite{tong2014probability}.
A stronger asymptotic bound on F.W.E.R. of Bonferroni's procedure has been provided in \cite{das2019bound}. \\
However, we do not want to restrict ourselves in asymptotic results. We have studied change in the behaviour of F.W.E.R. of these methods from small to large no. of observations. We have also studied the behaviour of F.D.R. for the Benjamini-Hodgeberg procedure as a function of correlation by simulation study. This paper is mainly targeted to exhibit the behaviour of the type-I errors by simulation study and provide an insight about how conservative the methods become in presence of correlation.

\section{ Description of the setup } 

Let $X_1,X_2, \dots $ be a sequence of random variables and we have a sequence of null hypotheses 
$$ H_{0i} : X_i \sim N(0,1) \: \: \: i=1,2,..... $$
 
The observations are not independent. For our study, we have considered equicorrelated setup. (i.e. $Corr (X_i, X_j) = \rho \: \: \forall \: i \neq j $). Suppose we have $n$ such null hypotheses ($H_{01} , \dots , H_{0n}$) and individual tests are one-sided.(i.e. $i$-th test rejects $H_{0i}$ if $X_i > c $ for some cut-off $c$). For the simultaneous $n$ tests define the family wise error rate (F.W.E.R.) in the following way. $$ F.W.E.R. = P( \bigcup\limits_{i=1}^{n} \{ X_i > c \} ) $$  

And the false discovery rate (F.D.R.) is defined in the following way. 

$$ FDR =E [  \frac{V}{R} I_{ R> 0 } ]  $$
Here $V$ represents number of false rejections and $R$ represents the number of rejections. \\
\vspace{0.1 in} 

It is interesting to note that, when all the null-hypotheses are true then $F.W.E.R. = F.D.R.$. 

 Our aim is to study this F.W.E.R. of Bonferroni's procedure and F.D.R. of Benjamini-Hodgeberg procedure as the correlation ($\rho$) varies in $[0,1]$. \\
 
Bonferroni's method sets a single cut-off for all the hypotheses. It rejects the $i$-th null hypothesis ($H_{0i}$) if $X_i > z_{ 1 - \frac{ \alpha}{n} }$. (Where $z_{ 1 - \frac{ \alpha}{n} }$ is the $(1- \frac{ \alpha}{n})-$th quantile of the standard normal distribution.) \\ 

\vspace{0.1 in} 

Benjamini-Hodgeberg (B.H.) procedure is a step-up procedure based on the P-values of individual tests. Suppose $P_1, P_2, \dots , P_n$ be the p-values and   $P_{(1) } \leq P_{(2) } \leq \dots \leq P_{(n) } $ be the ordered p-values and $H_{(0i)}$ be the hypotheses for which the p-value is $P_{(i)}$. Under the model described above, $P_i = \Phi ( -X_i) \: \: ( i=1,2, \dots , n)$. Then B.H. procedure rejects the set of $\{ H_{(0i)} \}$ such that, $\{ 1 \leq i \leq k | k = \max \{ j | P_{ (j) } \leq \frac{j \alpha}{n} \}$ and accepts every null hypotheses if $P_{j} \geq \frac{ j \alpha}{n} \: \: \: \forall j $.

\section{Simulation study} 
\subsection{Bonferroni's procedure} 
In our simulation study, we have generated $n$ observations from $N(0,1)$ distribution in such a way that there is a fixed correlation ($\rho$) between any two observations. It is important to note that, all the null hypotheses ($H_{0i}$) are true in our simulation study. We are only interested to study the behaviour of the type-I error here. Different combinations of $(n, \rho)$ are considered and they are studied at $\alpha=0.1,0.05,0.01$(Which are the most common levels of significance used for the purpose of hypothesis testing). 
\\
All the FWERs are estimated based on $10^5$ replications for Bonferroni's procedure at level of significance $\alpha $. In each plot the reference line $L(\rho) = \alpha (1- \rho) $ is attached in order to compare how much the F.W.E.R. curves lies below or above this line. For the Benjamini-Hodgeberg procedure, the F.D.R.s are also estimated based on $10^5$ replications for every $\alpha$.\\

The plots of F.W.E.R. and F.D.R. as a function of correlation ($\rho$) are shown below. 

\newpage
\begin{figure}[H]
\caption{Plots of FWER for various values of N ($\alpha = 0.01$) }
\label{fig:0.011}
\begin{center}
\includegraphics[width = 7.5 cm]{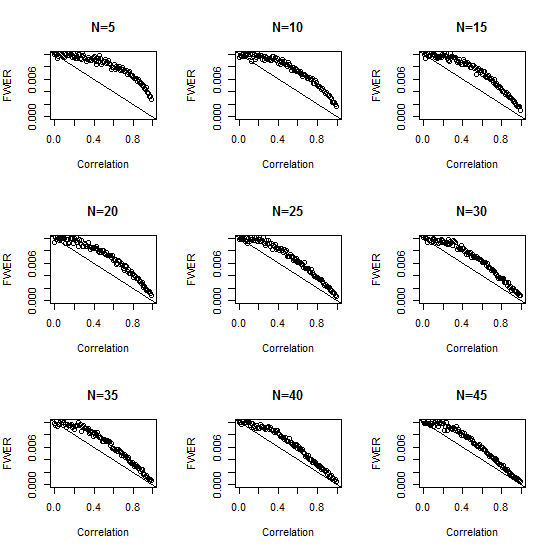} \\ 
\includegraphics[width = 7.5 cm ]{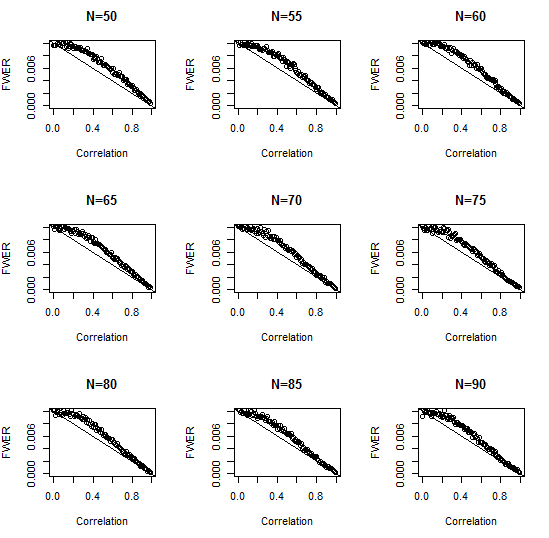} \\ 
\includegraphics[width=7.5  cm ]{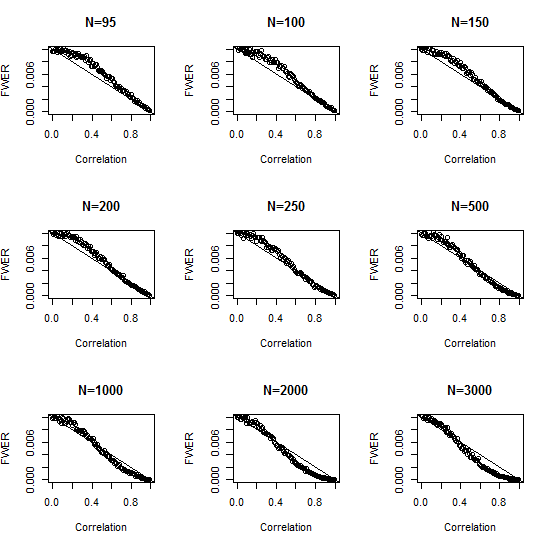} \\ 
\includegraphics[width = 7.5 cm]{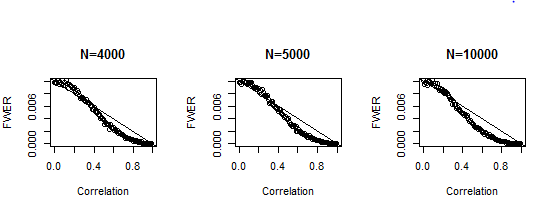}
\end{center} 
\end{figure}

\newpage

\begin{figure}[H]
\caption{Plots of FWER for various values of N ($\alpha = 0.05$) }
\label{fig:0.051}
\begin{center}
\includegraphics[width=7.5cm]{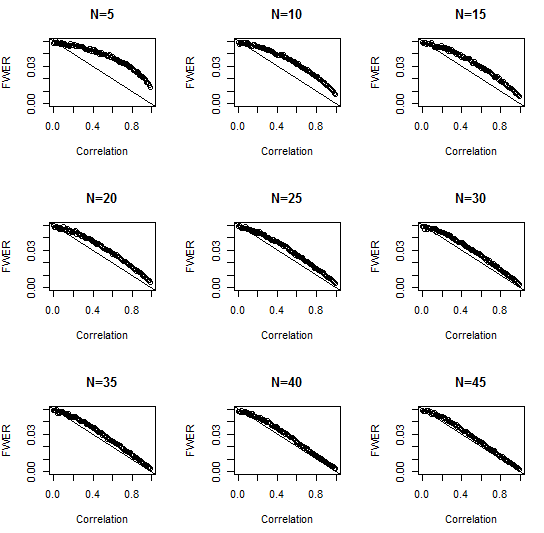} \\ 
\includegraphics[width=7.5cm]{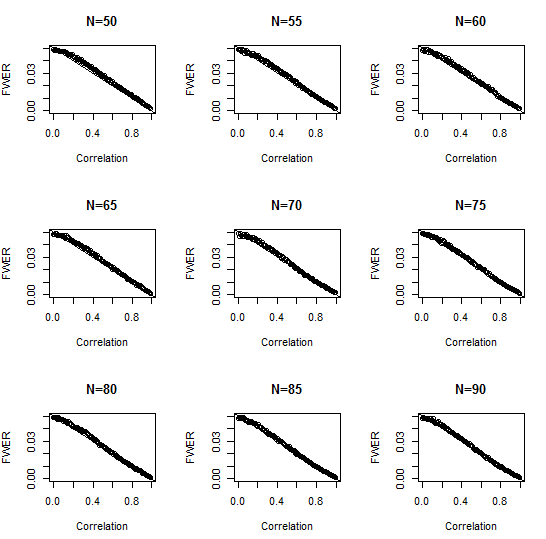} \\ 
\includegraphics[width=7.5cm]{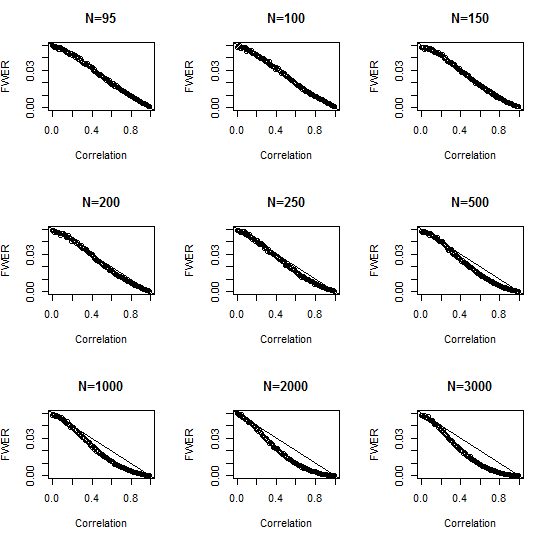} \\ 
\includegraphics[width=7.5cm]{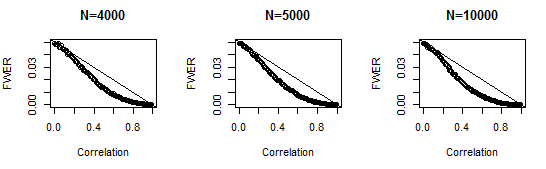} 
\end{center} 
\end{figure}

\newpage

\begin{figure}[H]
\caption{Plots of FWER for various values of N ($\alpha = 0.1$) }
\label{fig:0.11}
\begin{center} 
\includegraphics[width= 7.5 cm]{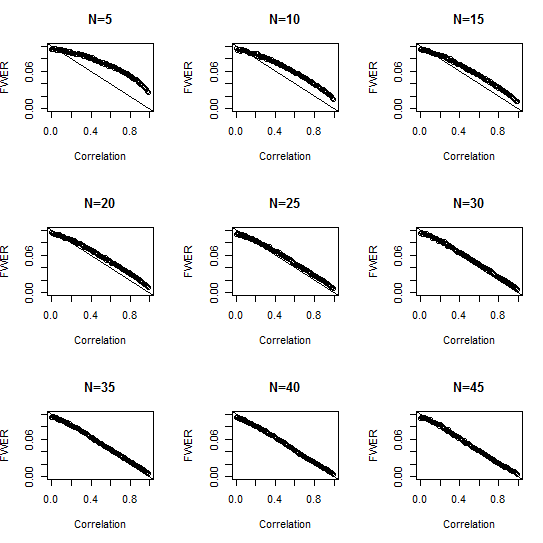} \\ 
\includegraphics[width= 7.5 cm]{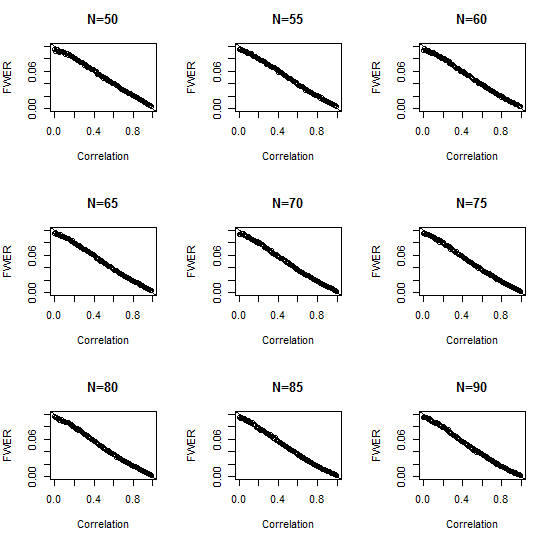} \\ 
\includegraphics[width=7.5 cm]{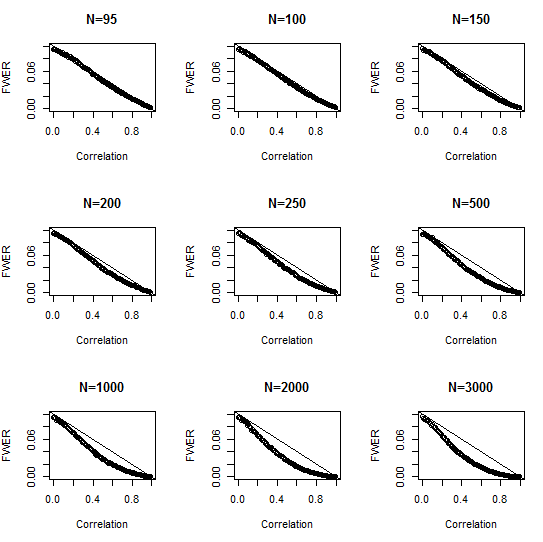} \\ 
\includegraphics[width=7.5 cm]{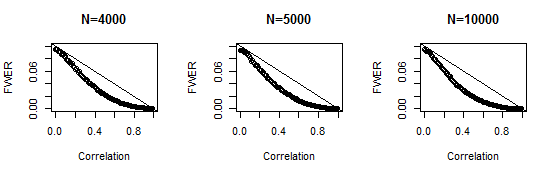}
\end{center}
\end{figure}

\newpage

\begin{figure}[H]
\caption{Plots of F.D.R. of B.H. procedure for various values of N ($\alpha = 0.01$) }
\label{fig:BH0.011} 
\begin{center} 
\includegraphics[width= 7.5 cm]{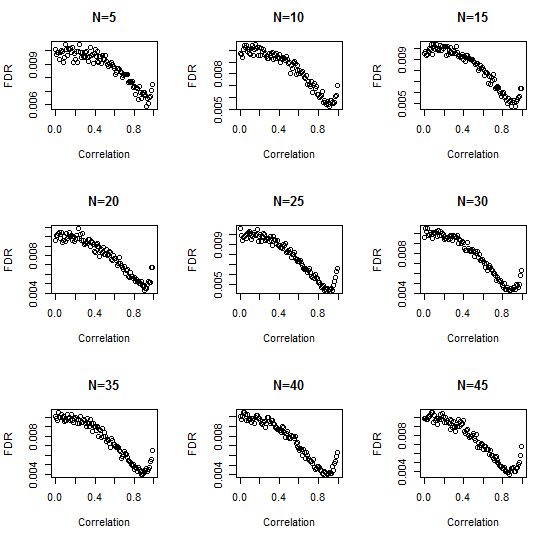} \\ 
\includegraphics[width= 7.5 cm]{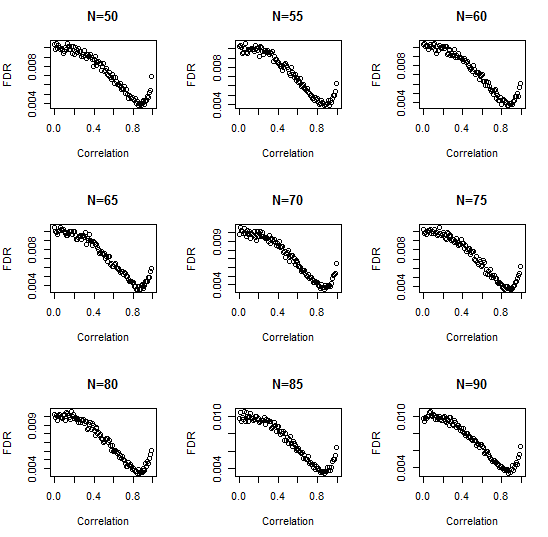} \\ 
\includegraphics[width= 7.5 cm]{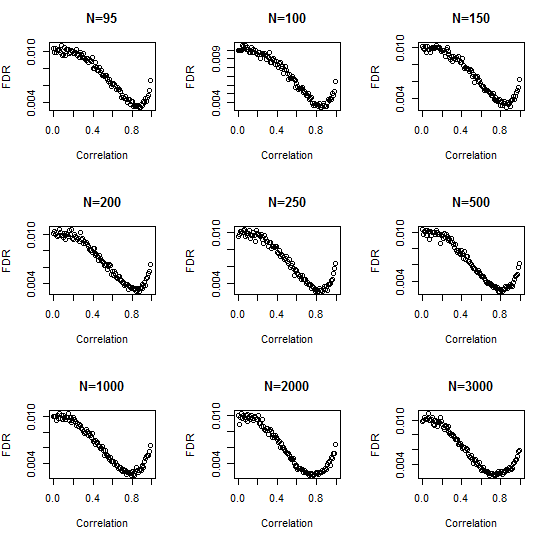} \\ 
\includegraphics[width= 7.5 cm]{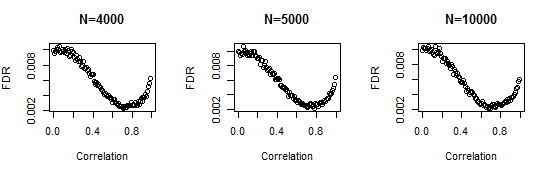} 
\end{center} 
\end{figure}

\newpage

\begin{figure}[H]
\caption{Plots of F.D.R. of B.H. procedure for various values of N ($\alpha = 0.05$) }
\label{fig:BH0.051}
\begin{center} 
\includegraphics[width= 7.5 cm]{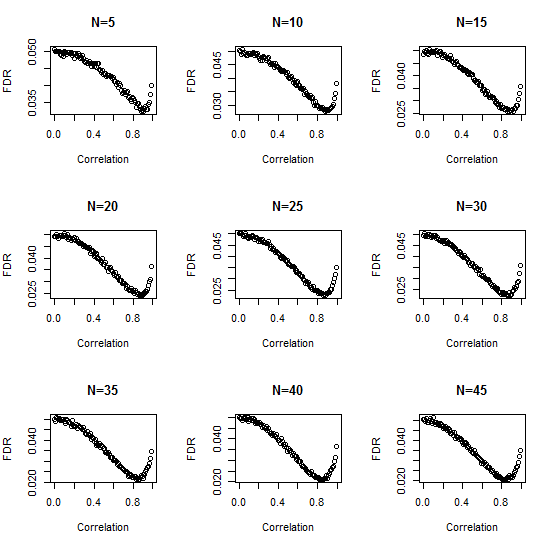} \\ 
\includegraphics[width= 7.5 cm]{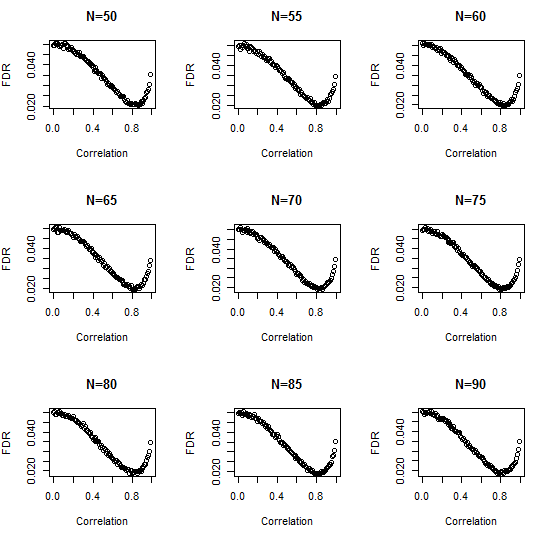} \\ 
\includegraphics[width= 7.5 cm]{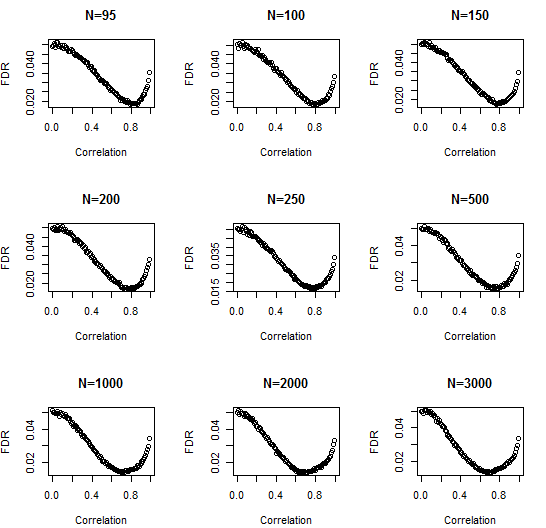} \\ 
\includegraphics[width= 7.5 cm]{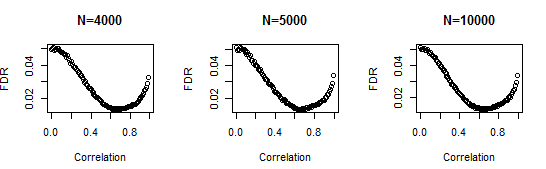} 
\end{center} 

\end{figure}

\newpage

\begin{figure}[H]
\caption{Plots of F.D.R. of B.H. procedure for various values of N ($\alpha = 0.1$) }
\label{fig:BH0.11} 
\begin{center} 
\includegraphics[width= 7.5 cm]{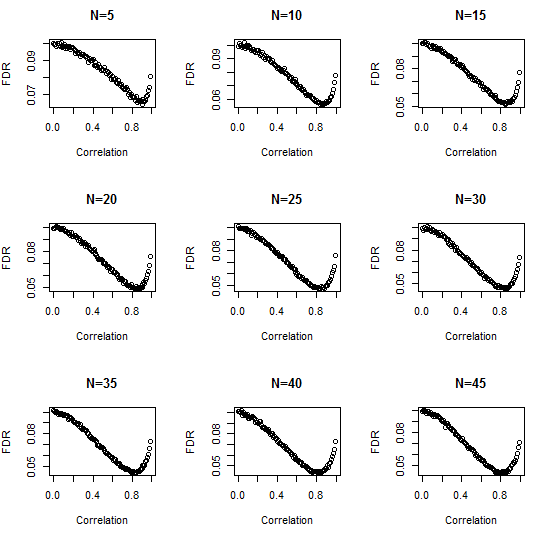} \\ 
\includegraphics[width= 7.5 cm]{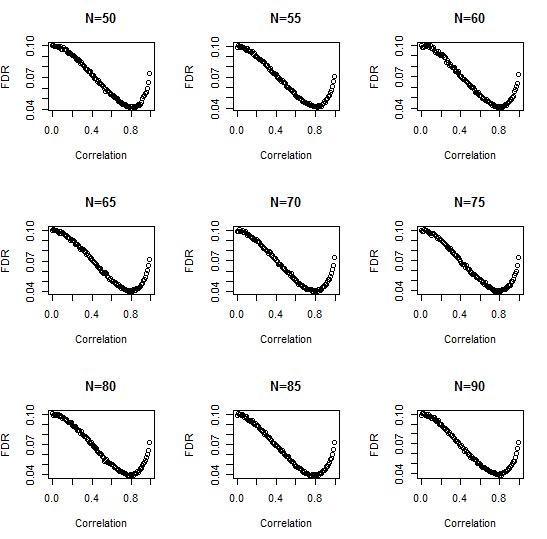} \\ 
\includegraphics[width= 7.5 cm]{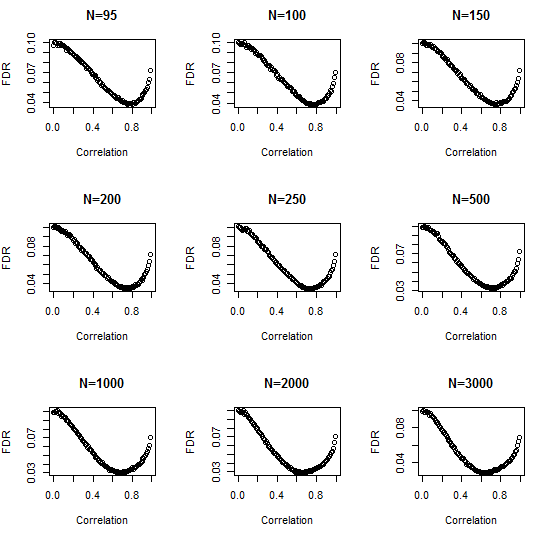} \\ 
\includegraphics[width= 7.5 cm]{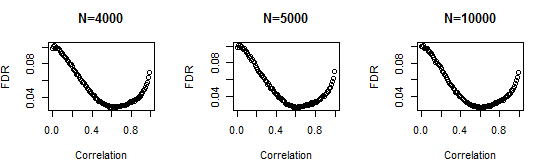} 
\end{center} 

\end{figure}

\newpage 

\subsection{Observations and discussion} 
There is a common pattern in the F.W.E.R. plots for all three $\alpha$'s taken in our simulation study. The F.W.E.R. for Bonferroni's procedure is a concave function for small $n$ and then gradually it changes it's nature. The plots confirm the claim of \cite{das2019bound} that, F.W.E.R. is asymptotically a convex function in $\rho$. But how quickly the concavity changes to convexity possibly depends on the value of $\alpha$. Observe that, in
figure~\ref{fig:0.11} (for $\alpha =0.1$ ) the nature changes very quickly while in figure~\ref{fig:0.011} ($\alpha = 0.01 ) $ the convergence to convexity occurs slowly. The reference line $L(\rho)= \alpha (1 - \rho) $ confirms the bound on F.W.E.R. for the asymptotic case (Shown in (\cite{das2019bound})).\\

We have taken all the $n$ null hypotheses to be true in our computation. In this setup, the F.W.E.R. calculation for Bonferroni's procedure and Holm's procedure is exactly similar. So, this simulation study also describes the F.W.E.R. of Holm's procedure. Both of them become conservative when correlation is present. If the correlation can be estimated in a consistent manner (let $\hat{\rho} $ be that estimate) , then to provide control at level $\alpha $ , the procedure should use $ \frac{ \alpha}{ 1 - \hat{ \rho} } $ as the corrected level of significance. This will provide a significant gain in  power for these methods. \\

The second part of the simulation study mainly focuses on the F.D.R. of Benjamini-Hodgeberg procedure. It is clear from the plots that this procedure is not as conservative as the F.W.E.R. controlling procedures. However, the F.D.R. tends to decrease at first and then increase as $\rho$ increases. But in all the cases, the F.D.R.'s are well below the desired level of significance ($\alpha$). It is a more challenging task to provide an asymptotic upper bound on the F.D.R. of the Benjamini-Hodgeberg procedure which enables us to provide a correlation correction in this setup. Much research is  needed in order to develop a stronger upper bound on these type-I error rates under general correlation structure.

\bibstyle{spbasic}

\bibliography{references.bib}

\end{document}